\documentclass[prl,twocolumn,floatfix]{revtex4}
\usepackage{bm}
\usepackage{graphicx}
\usepackage{amsmath}

\newcommand{\rsub}[2]{\mathbf{r}_{#1}^{#2}}
\newcommand{\Mset}{\{ M_i \}}
\newcommand{\Msetmax}{\Mset_\text{max}}
\newcommand{\mset}{\{ m_i \}}
\newcommand{\msetmax}{\mset_\text{max}}
\newcommand{\avib}{a_\text{vib}}
\newcommand{\pderiv}[2]{\frac{\partial #1}{\partial #2}}
\newcommand{\erfi}{\text{erfi}}
\newcommand{\sfrac}[2]{{\textstyle\frac{#1}{#2}}}

\begin{document}

\title{Saddles in the energy landscape: extensivity and thermodynamic formalism}

\author{M. Scott Shell}
\email[email:]{shell@princeton.edu}
\author{Pablo G. Debenedetti}
\email[email:]{pdebene@princeton.edu}
\author{Athanassios Z. Panagiotopoulos}
\email[email:]{azp@princeton.edu}
\affiliation{Department of Chemical Engineering\\Princeton University, Princeton, NJ 08544}
\date{May 28, 2003}

\begin{abstract}
We formally extend the energy landscape approach for the thermodynamics of liquids to account for saddle points.  By considering the extensive nature of macroscopic potential energies, we derive the scaling behavior of saddles with system size, as well as several approximations for the properties of low-order saddles (i.e., those with only a few unstable directions).  We then cast the canonical partition function in a saddle-explicit form and develop, for the first time, a rigorous energy landscape approach capable of reproducing trends observed in simulations, in particular the temperature dependence of the energy and fractional order of sampled saddles.
\end{abstract}

\maketitle

In recent years significant effort has been devoted to the study of supercooled liquids and their glasses \cite{glasses}.  An important aspect of  these technologically significant \cite{angell1} systems is the interplay between dynamic and thermodynamic processes, which is thought to play a key role in their kinetic slowdown and eventual falling out of equilibrium at the glass transition \cite{pgd1}.  The energy landscape formalism of Stillinger and Weber has been a useful tool in the theory of supercooled liquids \cite{sw1}.  In this description, a system's configuration space is partitioned into basins surrounding local energy minima.  Termed ``inherent structures,'' these minima correspond to mechanically stable particle packings and are described statistically by their depth: $\exp{\left[ N \sigma(\phi) \right]}d\phi$ gives the scaling of the number of distinct minima with per-particle potential energy (or depth) $\phi \pm d\phi /2$, where $N$ is the number of particles and $\sigma$ is called the basin enumeration function.  Here, ``distinct'' refers to minima differing by more than mere particle permutation.  This formalism permits a rigorous transformation of the canonical partition function:
\begin{equation}
Z \sim \int e^{ N \left[ \sigma(\phi) - \beta \phi - \beta \avib(\beta, \phi) \right] } d\phi
\label{eq:sw}
\end{equation}
where $\beta=1/k_B T$ and $\avib$, the vibrational free energy, is the per-particle free energy when the system is confined to an average basin of depth $\phi$.  For each temperature in the thermodynamic limit, the system samples basins of a well-defined energy $\phi^*$; deeper basins are accessed as the temperature decreases.  One identifies a configurational entropy, $N k_B \sigma(\phi^*)$, which is that part of the entropy due to the multiplicity of amorphous configurations explored by the system.  Good functionalities can be rationalized for $\sigma$ and $\avib$ (and measured in computer simulations) such that the partition function can be explicitly evaluated \cite{pgd2}; this approach is also useful for characterizing kinetic processes as it has been observed that the configurational entropy plays a key role in dynamics \cite{pgd1,speedy1}.

Our work aims to extend the energy landscape formalism to include a description of higher order stationary points, i.e., saddles in the landscape \cite{keyes1}.  This approach provides a natural connection with dynamics \cite{keyes2,wales1}, but is more intricate than minima alone because, in addition to their energy, saddles are also classified by their order\textemdash the number of directions with negative curvature.  In this Letter, we begin by deriving the extensivity properties of saddles and propose their corresponding enumeration function.  We then derive a saddle ``equipartition'' theorem and show the expected scaling behavior for low-order saddles.  Finally we give the appropriate form of the partition function in this formalism and demonstrate its utility for describing the behavior of supercooled liquids.

Our consideration of saddles relies on an extensive macroscopic potential energy.  That is, a single macroscopic system of $N$ particles can be effectively divided into $M \ll N$ equivalent smaller subsystems, between which boundary interactions are negligible compared to the total energy.  The number of particles in each subsystem is macroscopic, $N_S \gg 1$, but the number of subsystems is also large, $N_S \ll N$.  This condition is satisfied by most common types of molecular interaction (notable exceptions include molecules with long-range interactions or which are themselves macroscopic in size).  For a single-component system of structureless particles, the potential energy can then be written as
\begin{equation}
U(\mathbf{r}^N) \approx U(\rsub{(1)}{N_S}) + U(\rsub{(2)}{N_S}) + \ldots + U(\rsub{(M)}{N_S})
\label{eq:energysep}
\end{equation}
where $U$ is the potential energy function, $\rsub{}{N} \equiv \{\rsub{1}{},\rsub{2}{} \ldots \rsub{N}{}\}$ gives the positions of the particles, and $\rsub{(i)}{N_S}$ are the corresponding positions of the $N_S$ particles in subsystem $i$.  Here, any stationary point in the overall system can be viewed as a combination of stationary points in each subsystem.  It is relatively straightforward to determine the scaling behavior of minima \cite{stillinger1}: if the number of distinct minima in each subsystem is $g_0$, the total number due to their possible combinations is $g_0^M$, or, $\exp{[N \ln{(g_0)}/N_S]} \equiv \exp{[N \sigma_\infty]}$ \cite{com:function}.  Our notation for the density-dependent constant $\sigma_\infty$ indicates its correspondence with the total number of inherent structures (not of a particular energy); equivalently, $\sigma_\infty$ is the maximum value of the basin enumeration function.  

For saddles, one must consider their order $n$, defined as the number of negative eigenvalues in the Hessian matrix, $H_{ij} \equiv \frac{\partial^2 U}{\partial r_i \partial r_j}$.  Imposition of Eq.\ \ref{eq:energysep} gives rise to a Hessian which is reducible in each of the subsystems; therefore, the total number of negative eigenvalues is the sum of that for each of the subsystems.  In other words, the saddle order in the total system is the sum total of the orders of the subsystems.  We consider a particular distribution of the total saddle order $n$ among the subsystems, letting the values $M_i$ denote the number of subsystem saddles of order $i=0,1 \ldots d N_S$ ($d$ = dimensionality).  In this notation, the constraints are $\sum_{i} M_i = M$ and $\sum_{i} i \cdot M_i = n$.  For an overall saddle order $n$ and a particular distribution $\Mset$, the number of distinct saddles is
\begin{equation}
\Omega_n (\Mset) = ( M! / \prod_i M_i! )  \prod_i g_i^{M_i} 
\label{eq:omegaMset}
\end{equation}
where $g_i$ is the number of distinct saddles of order $i$ in a subsystem.  The total number of saddles, however, must be the sum of $\Omega_n(\Mset)$ for all possible distributions $\Mset$.  We will find that one particular distribution, $\Msetmax$, overwhelmingly dominates this sum.  First we switch to an ``intensive'' notation by introducing the following variables: $m_i \equiv M_i/M$, and $x \equiv n/d N$ is the overall fractional saddle order.  Insertion into Eq.\ \ref{eq:omegaMset} and application of Stirling's approximation yields
\begin{equation}
\Omega_x (M,\mset) = \left[ \prod_{i} \left( g_{i} \big/ m_i \right)^{m_i} \right]^{M} .
\label{eq:omegamset}
\end{equation}
Similarly, the constraint equations in intensive form become $\sum_{i} m_i = 1$ and $\sum_{i} (i/d N_S) \cdot m_i = x$.  These constraints and the terms inside the brackets in Eq.\ \ref{eq:omegamset} are all independent of the number of subsystems $M$.  As a result, the distribution $\msetmax$ which maximizes the term in brackets depends only on the overall fractional saddle order $x$.  In the thermodynamic limit, $M = N/N_S \to \infty$, this maximum term dominates the sum over all distributions $\mset$.  These considerations lead directly to the saddle scaling behavior:
\begin{eqnarray}
\Omega_x (N) & = & \left[ \prod_{i} \left( g_{i} \big/ m_i \right)^{m_i} \right]_\text{max}^{N/N_S} 
\nonumber \\
& \equiv & \exp{\left[ N \theta_\infty(x) \right]}
\label{eq:omegax}
\end{eqnarray}
where $\Omega_x$ gives the number of distinct saddles of fractional order $x$.  Here we have introduced the generalized function $\theta_\infty(x)$ which characterizes saddle scaling behavior and has the property $\theta_\infty(x \to 0) = \sigma_\infty$.  Notice that the relevant order parameter for saddles is their fractional order, such that $\exp{[N \theta_\infty(x)]}dx$ gives the number of saddles with fractional order $x \pm dx/2$.

Returning to the subsystem scenario, we now find the distribution $\msetmax$ which gives the dominant saddles in the system.  To do so, one maximizes the term in brackets in Eq.\ \ref{eq:omegamset} or equivalently, its logarithm.  Using Stirling's approximation, the distribution $\msetmax$ must satisfy
\begin{equation}
\max{\sum_{i} m_i \ln{g_i} - m_i \ln{m_i}}.
\label{eq:maxterm2}
\end{equation}
By accounting for the constraints with the usual Lagrange multipliers and using Eq.\ \ref{eq:omegax} for $g_i = \exp{[N_S \theta_\infty(i/d N_S)]}$, the following form arises when Eq.\ \ref{eq:maxterm2} is evaluated:
\begin{equation}
m_i = \frac{ \exp{[N_S \theta_\infty(z) - \gamma N_S z]} }{ \sum_{i'} \exp{[N_S \theta_\infty(z') - \gamma N_S z']} }
\label{eq:lagrange}
\end{equation}
where $z \equiv i /d N_S$ is the fractional order of a subsystem.  Here $\gamma$ is a Lagrange multiplier ensuring the constraint $\sum_i z \cdot m_i = x$.  Because the terms in the exponential grow with the size of the subsystems, which are themselves macroscopic, $m_i$ is essentially zero for all $i$ except one, $i_\text{max}$.  With this simplification, the constraint yields $i_\text{max} = n/M$, or $z_\text{max} = x$.  In other words, the total saddle order is distributed across the subsystems such that their fractional saddle order is equivalent to each other and to the overall fractional order.  This is in effect an ``equipartition'' of saddle order across the geometry of the system.

Such equipartition has important consequences for low-order saddles; it implies that the majority of such saddles are built from a collection of localized first-order saddles.  In this sense, each direction of negative curvature in the potential energy corresponds to an elementary saddle ``defect'' in an inherent structure.  This observation can be used to determine the approximate behavior of $\Omega_x$ for small values of $x$.  We assume that $\xi d N$ non-interacting, first-order defects are possible for any inherent structure.  The constant $\xi$, for example, may attain a value close to $1/d$ if each molecule in the system can move independently of the others so as to form a saddle.  This yields
\begin{equation}
\Omega_{x \ll 1} \sim \exp{[N\sigma_\infty]} \times (\xi d N)! \big/ (\xi d N - x d N)! (x d N)!
\label{eq:omegalowx}
\end{equation}
or, taking the logarithm and applying Stirling's approximation,
\begin{equation}
\theta_\infty(x \ll 1) \approx \sigma_\infty - \sfrac{x}{\xi} \ln{\sfrac{x}{\xi} }
- \left(1 - \sfrac{x}{\xi} \right) \ln{\left( 1 - \sfrac{x}{\xi} \right)}.
\label{eq:thetalowx}
\end{equation}
We can also state for a system containing non-interacting first-order saddles that the average saddle energy depends linearly on order:
\begin{equation}
\overline{\Phi}_n = \overline{\Phi}_0 + n  \left( \overline{\Phi}_1 - \overline{\Phi}_0 \right)
\label{eq:saddleenergy}
\end{equation}
where $\overline{\Phi}_n$ is the average energy of an $n$th order saddle.  This trend has been discussed previously in theoretical work \cite{keyes1} and has been found in simulation studies to be appropriate \cite{sciortino1,broderix1}.  One must bear in mind, however, that both Eq.\ \ref{eq:saddleenergy} and the equipartition of saddles apply to the entire ensemble of stationary points, of which only a minute fraction are sampled by the system in equilibrium at low temperature.  A system may require, for example, the cooperative movement of many molecules in order to reach nearby saddles, which may cause low energy stationary points to be of higher order than expected.  The low-$T$ persistence of a linear relationship between saddle order and energy observed in simulation therefore suggests that the first-order defect scenario persists even for low-energy saddles; cooperativity then arises because the direction of negative potential energy curvature about these points is a superposition of several molecules' atomic coordinates.

The considerations so far have categorized saddles only by their fractional order.  Following the approach used for inherent structures \cite{sw1}, one might extend this description to potential energy as an additional order parameter.  We therefore introduce the saddle enumeration function, $\theta(\phi,x)$, for which the expression $\exp{[N\theta(\phi,x)]}d\phi dx$ is proportional to the number of saddles with potential energy per-particle $\phi \pm d\phi /2$ and of fractional order $x \pm dx/2$.  (The basin enumeration function is retrieved from $\theta(\phi,x = 0)$.)  This extension allows a meaningful casting of the canonical partition function in which configuration space is divided into ``basins'' surrounding saddle points \cite{com:basins}:
\begin{equation}
Z \sim \int_{x=0}^{x=1} \int_{\phi_\text{min}}^{\phi_\text{max}}
e^{ N \left[ \theta(\phi,x) - \beta \phi - \beta \avib(\beta, \phi, x) \right] } d\phi dx.
\label{eq:partitionfunction}
\end{equation}
In this equation, $\avib$ is the vibrational free energy around a stationary point of energy $\phi$ and fractional order $x$; formally it is given by
\begin{equation}
e^{N\avib(\beta,\phi,x)} \equiv \Lambda^{-d N} \left< \int_{\mathbf{\Gamma}_k} 
e^{ \beta \left[ U(\mathbf{r}^N) - N \phi \right] } d\mathbf{r}^N \right>_{\phi, x}
\label{eq:avib}
\end{equation}
where $\Lambda$ is the thermal deBroglie wavelength, the average is restricted to saddle points of energy $\phi$ and order $x$, and the integral for a particular saddle $k$ is performed over its associated configuration space $\mathbf{\Gamma}_k$ \cite{com:basins}.  In the large system limit, the integral in Eq.\ \ref{eq:partitionfunction} will be dominated by the maximum value in the exponential, and the conditions for equilibrium can be written as
\begin{eqnarray}
\pderiv{\theta}{\phi} = \beta \left(1+\pderiv{\avib}{\phi}\right) ;
\pderiv{\theta}{x} = \beta \pderiv{\avib}{x} .
\label{eq:equilib}
\end{eqnarray}
The simultaneous solution to these equations provides the average saddle energy $\phi^*$ and order $x^*$ sampled by the system at specified temperature.  The total Helmholtz free energy is then $A/N = \avib(\beta,\phi^*,x^*) + \phi^* - k_B T \theta(\phi^*,x^*)$.  One can identify from this equation a per-particle saddle entropy, $k_B \theta$, which converges on the conventional Stillinger-Weber configurational entropy at very low temperatures when the system spends most of its time near minima and $x^*$ asymptotes to zero.  This point of view sets the stage for a more rigorous connection with dynamics, for which $x^*$ contains pertinent information.

Knowledge of the functions $\theta$ and $\avib$ provides all the thermodynamic details of the system.  We now show that a number of reasonable assumptions about their functional form results in a physically realistic and insightful picture.  Our analysis addresses results for several well-studied glass-forming systems for which numerical data exist \cite{kobandersen1,sastry1,sciortino1,broderix1,wales1}.  First, we assume the vibrational free energy to be independent of saddle energy, $\avib(\beta,\phi,x) \approx \avib(\beta,x)$.  This is rigorously true for minima at absolute zero, but remains a working simplification in our analysis.  Furthermore, we model the vibrational free energy in the classical harmonic approximation \cite{hill1}:
\begin{equation}
e^{-N \beta \avib}  \approx  \Lambda^{-d N} 
\left[ \int_{-l}^{l} e^{-\beta\alpha_S  r^2} dr \right]^{d N -n}
\left[ \int_{-l}^{l} e^{\beta\alpha_U  r^2} dr \right]^{n}
\label{eq:zvibapprox}
\end{equation}
\begin{eqnarray}
\beta \avib & \approx & d \ln{\left(T_S^{1-x}T_U^x/T\right)} - x \ln{\erfi\sqrt{\beta \alpha_U l^2}}
\nonumber \\
& \approx & d \ln{\left(T_S/T \right)} - x \left[ C \beta - \ln{(\pi C \beta)}/2 \right]
\label{eq:avibapprox}
\end{eqnarray}
where $\alpha_S,\alpha_U \equiv \frac{1}{2} \left| \frac{\partial^2 U}{\partial r^2} \right|$ are half the average curvatures of stable and unstable modes, respectively; $T_S$ and $T_U$ are the corresponding Einstein temperatures \cite{hill1}; $l$ is a length scale characteristic of the saddle's associated configuration space volume \cite{wales2}; and $\erfi$ is the imaginary error function.  In the last line, we assume $\alpha_S = \alpha_U$ and use the asymptotic expansion of the error function, $\erfi(x) \to \exp{[x^2]}/x \sqrt{\pi}$, for the low $T$ limit.  $C=\alpha_U l^2$ is a lumped constant.  In general, the harmonic approximation is increasingly valid for stable modes at low temperatures, but we have made liberal use of its application to the unstable modes, especially in that the final $\avib$ depends nontrivially on the length scale $l$ characterizing the  $dN$-dimensional integral.  Nonetheless, this approximation provides a starting point for analysis, and we leave investigation of more accurate forms to future work.

For the saddle enumeration function, we assume a Gaussian form in energy, consistent with previous simulation studies \cite{heuer1,sastry1,wales1,sconf2}:
\begin{equation}
\theta(\phi,x) = \theta_\infty(x) \left[1 - \left(\phi - \overline{\phi}(x) \right)^2 / \Delta^2 \right]
\label{eq:thetaapprox}
\end{equation}
where $\overline{\phi}$ is the average energy of saddles of fractional order $x$ and $\Delta$ is their characteristic energy range.  For the dependence of the parameters in this expression on fractional saddle order, we use Eq.\ \ref{eq:thetalowx} for $\theta_\infty$, implement the linear relationship in Eq.\ \ref{eq:saddleenergy} for $\overline{\phi}$ such that $\overline{\phi}(x)=\overline{\phi}_0 + \delta \cdot x$, and assume $\Delta$ to be roughly constant.  

\begin{figure}
\includegraphics[width=3.375 in]{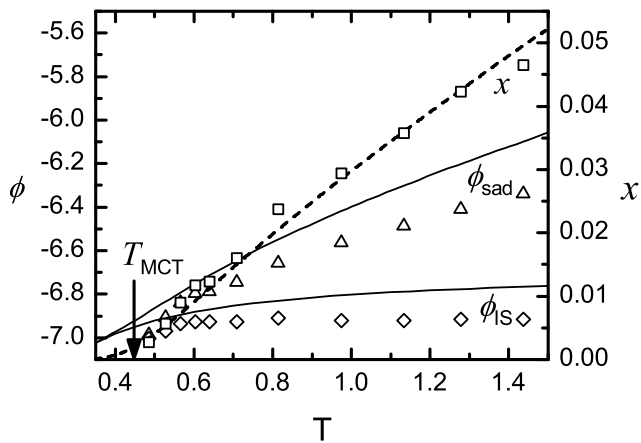}
\caption{Plot of the equilibrium fractional saddle order ($x$), and average saddle ($\phi_\text{saddle}$) and inherent structure ($\phi_\text{IS}$ ) energies as a function of temperature. The squares, triangles, and diamonds are the respective results for a modified Lennard-Jones system \cite{sciortino1}.  $T_\text{MCT} \approx 0.435$ is the mode-coupling temperature \cite{kobandersen1}.}
\label{fig:theory}
\end{figure}

The usefulness of this approach can be seen in the predictions of the theory.  Such an analysis is possible by solving Eq.\ \ref{eq:equilib} with the simplified expressions for the vibrational free energy and the saddle enumeration function.  We choose representative parameters based on previous simulations \cite{com:params}, though two have no precedent in the literature, which we instead choose so that $x^*(T)$ matches the simulation result in Ref.\ \onlinecite{sciortino1}.  Our results are shown in Fig.\ \ref{fig:theory}.  Overall, the theory and assumptions capture the essential behavior observed in simulations (the general shape and relationship of the curves in Fig.\ \ref{fig:theory}) and are a promising starting point for further refinement of the approach.  We note that our parameters are chosen among several similar model systems, and quantitative agreement in Fig.\ \ref{fig:theory} might be possible if a complete dataset for any one system were available.

In summary, we have presented a thermodynamic formalism which includes higher order stationary points in the energy landscape.  Through a reformulation of the canonical partition function and by using several physically-motivated simplifications, we show that this formalism captures important trends in the behavior of low-$T$ glass-forming materials.  Future work will investigate the relationship suggested by this approach between liquid kinetics and thermodynamics.

We gratefully acknowledge the support of the Fannie and John Hertz Foundation and of the Dept.\ of Energy, Division of Chemical Sciences, Geosciences, and Biosciences, Office of Basic Energy Science (grants DE-FG02-87ER13714 to PGD and DE-FG02-01ER15121 to AZP.)


\begin{thebibliography}{23}
\bibitem{com:function} Equivalently, the function $f$ describing the scaling behavior must satisfy $f(N_S)^{M}=f(M N_S)$.
\bibitem{com:basins} A stationary point's basin might be defined as those points in configuration space which are closer to it than any other (as defined by the $d N$-dimensional distance).
\bibitem{com:params} The following parameters are employed, all in reduced units: $\sigma_\infty$=0.933, $\overline{\phi}_0 = -6.683$, $\delta=13.3$, $\Delta=0.472$, $\xi=2.22$, $C=13.34$.  These values are taken from Refs.\ \onlinecite{sastry1} and \onlinecite{sciortino1}, except for the last two which are fit.
\end{thebibliography}
\end{document}